# An Optimal Coding Strategy for the Binary Multi-Way Relay Channel

Lawrence Ong, Sarah J. Johnson, and Christopher M. Kellett

**Abstract**

We derive the capacity of the binary multi-way relay channel, in which multiple users exchange messages at a common rate through a relay. The capacity is achieved using a novel functional-decode-forward coding strategy. In the functional-decode-forward coding strategy, the relay decodes functions of the users' messages without needing to decode individual messages. The functions to be decoded by the relay are defined such that when the relay broadcasts the functions back to the users, every user is able to decode the messages of all other users.

## 1 Introduction

We derive the *common-rate* capacity of the binary multi-way relay channel (MWRC). The MWRC is a multicast network where the users exchange full information with one another. As there is no direct connection among the users, communication is done via a relay (e.g., see Fig. 1). We consider the case where every user sends independent messages at the same (common) rate to all other users. We will show that our proposed *functional-decode-forward* coding strategy achieves the common-rate capacity of the binary MWRC for any number of users and for all noise levels.

The MWRC is an extension of the two-way relay channel where two users exchange data via a relay (e.g., see [1, 2]). The Gaussian MWRC has been recently studied by Gündüz *et al.* [3], where three achievable common-rate regions have been derived using the complete-decode-forward[1], compress-forward, and amplify-forward coding strategies respectively. However, none of these three strategies achieve the common-rate capacity in general. In this paper, we consider a simpler binary MWRC to gain insights into optimal coding strategies for the general MWRC channel.

In the functional-decode-forward coding strategy, the relay only needs to decode functions of the users' messages, compared to decoding all users' messages in the complete-decode-forward coding strategy. The functions must be defined such that any user can decode other users' messages from the functions and its own message. Furthermore, in functional-decode-forward, noise in the *uplink* (the channel from the users to the relay) is removed at the relay, while in compress-forward and amplify-forward, the uplink noise propagates to the *downlink* (the channel from the relay to the users).

---

[1]We modified the strategy name "decode-and-forward" used in [3] to distinguish this coding strategy and our proposed functional-decode-forward coding strategy.

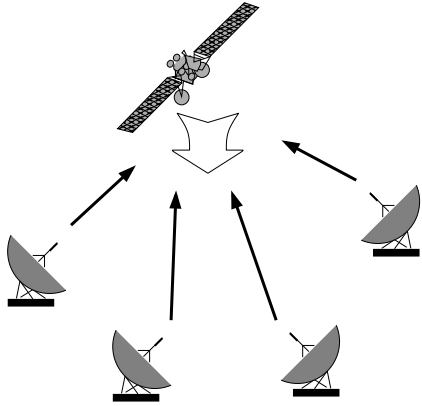

Figure 1: An example of a multi-way relay network, where stations exchange information via a satellite

We state the channel model in Sec. 2 and derive an upper bound to the common-rate capacity in Sec. 3. We then define the functional-decode-forward coding strategy proper and show that it achieves the capacity upper bound in Sec. 4. Section 5 concludes the paper.

## 2 Channel Model

We define the $L$-user binary MWRC as follows:

- Nodes 1, 2, ..., $L$ are the users, and node 0 is the relay,

- The channel input from node $i$ is denoted by $X_i \in \{0,1\}$, and the channel output received by node $i$, $Y_i \in \{0,1\}$, $\forall i \in [0, L]$.

- The uplink is

$$Y_0 = X_1 \oplus X_2 \oplus \cdots \oplus X_L \oplus E_0 = \bigoplus_{1 \leq i \leq L} X_i \oplus E_0, \qquad (1)$$

where $\oplus$ is defined as $\sum$ in modulo-2.

- The downlink consists of

$$Y_i = X_0 \oplus E_i, \quad i = 1, 2, \ldots, L. \qquad (2)$$

Here, the channel noise $E_i \in \{0,1\}$, $\forall i \in [0, L]$, are independent. $\Pr\{E_i = 1\} = \rho_i$ is commonly known as the cross-over probability of the binary-symmetric channel.

Each user $i$ encodes its message $W_i$ into a length $n$ codeword $\boldsymbol{X}_i = (X_i[1], X_i[2], \ldots, X_i[n])$, and transmits it to the relay. We consider the *restricted* MWRC in the sense that the transmit signals of each user are functions of its messages and are not functions of its received symbols. The relay itself has no data to send, and its transmit signal at time $t$, $X_0[t]$, can only depend on its previously received signals $\{Y_0[\ell] : 1 \leq \ell \leq t-1\}$. User $i$ attempts to decode all other users' messages after $n$ channel uses, i.e., from $\boldsymbol{Y}_i = (Y_i[1], Y_i[2], \ldots, Y_i[n])$.

We consider the *symmetric* case where the users' messages each have $nR$ bits. We say that the common rate $R$ is *achievable* if the probability that any node $i \in [1, L]$ wrongly decodes any message $W_j$, $j \in [1, L] \setminus \{i\}$, can be made arbitrarily small. The common-rate capacity $C$ is defined as the supremum of all achievable rates.

## 3  An Upper Bound to The Common-Rate Capacity

An upper bound on the common-rate capacity is given in the following theorem.

**Theorem 1.** *The common-rate capacity of the binary MWRC is upper-bounded by*

$$C \leq \min_{0 \leq i \leq L} \left\{ \frac{1 - H(\rho_i)}{L - 1} \right\}, \tag{3}$$

*where* $H(\rho_i) \triangleq -\rho_i \log_2(\rho_i) - (1 - \rho_i) \log_2(1 - \rho_i) = H(E_i)$.

*Proof of Theorem 1.* Consider a network of $m$ nodes, in which node $i$ sends information at the rate $R_{i,j}$ to node $j$. If the set of rates $\{R_{i,j}\}$ are achievable, there exists some joint probability distribution $p(x_1, x_2, \ldots, x_m)$ such that [4, page 589 (Theorem 15.10.1)]

$$\sum_{i \in \mathcal{S}, j \in \mathcal{S}^c} R_{i,j} \leq I(X_\mathcal{S}; Y_{\mathcal{S}^c} | X_{\mathcal{S}^c}), \tag{4}$$

for all $\mathcal{S} \subset \{1, 2, \ldots, m\}$. Here $X_\mathcal{S} = \{X_i : i \in \mathcal{S}\}$, and $\mathcal{S}^c = \{1, 2, \ldots, m\} \setminus \mathcal{S}$. This upper bound is often called the cut-set bound. A cut-set bound of a network is the maximum rate that information can be transferred across a *cut* separating two disjoint sets of nodes, assuming that all nodes on each side of the cut can fully cooperate.

Now, we apply the cut-set bound to the MWRC. First, we consider the cut separating $\mathcal{S} = \{1, 2, \ldots, i-1, i+1, \ldots, L\}$ for some $i \in [1, L]$, and $\mathcal{S}^c = \{0, i\}$. The total information flow from $\mathcal{S}$ to $\mathcal{S}^c$ is $(W_1, W_2, \ldots, W_{i-1}, W_{i+1}, \ldots, W_L)$ with the total common rate $(L-1)R$. We have the following rate constraint on $R$, for all $i \in [1, L]$:

$$(L-1)R \leq \left[ H(Y_0, Y_i | X_0, X_i) - H(Y_0, Y_i | X_{[0,L]}) \right] \tag{5a}$$

$$= H\left( \bigoplus_{i \in \mathcal{S}} X_i \oplus E_0, E_i \right) - H(E_0, E_i) \tag{5b}$$

$$= H\left( \bigoplus_{i \in \mathcal{S}} X_i \oplus E_0 \right) - H(E_0), \tag{5c}$$

where (5c) is because $\left( \bigoplus_{i \in \mathcal{S}} X_i \oplus E_0 \right)$ and $E_i$ are statistically independent, so are $E_0$ and $E_i$.

Now, we consider the cut separating $\mathcal{S} = \{0, 1, 2, \ldots, i-1, i+1, \ldots, L\}$ for some $i \in [1, L]$, and $\mathcal{S}^c = \{i\}$. The total information flow from $\mathcal{S}$ to $\mathcal{S}^c$ is again $(W_1, W_2, \ldots, W_{i-1}, W_{i+1}, \ldots, W_L)$ with the total common rate $(L-1)R$. We have the following rate constraint on $R$, for all $i \in [1, L]$.

$$(L-1)R = \left[ H(Y_i | X_i) - H(Y_i | X_{[0,L]}) \right]. \tag{6a}$$

$$= H(X_0 \oplus E_i) - H(E_i). \tag{6b}$$

The common rate $R$ must be bounded by the two constraints (5c) and (6b) for all $i$ and for some $p(x_0, x_1, \ldots, x_L)$. For any binary random variable $X$, its maximum entropy $H(X)$ is one and is attained by the uniform distribution $p_U(x)$. So, choosing the independent and uniform distribution $p(x_0, x_1, \ldots, x_L) = p_U(x_0) p_U(x_1) \cdots p_U(x_L)$ simultaneously maximizes (5c) and (6b) for all $i \in [0, L]$. Thus, we have Theorem 1. □

## 4 Functional-Decode-Forward

The concept of functional-decode-forward was first proposed for the two-way relay channel, i.e., $L = 2$ [5–7]. If nodes 1 and 2 transmit linear codes, the relay receives a noisy version of $\boldsymbol{X}_{1,2} = \boldsymbol{X}_1 \oplus \boldsymbol{X}_2$, which is another codeword. With error-correcting codes, the relay can decode $\boldsymbol{X}_{1,2}$, and broadcast it back to users 1 and 2. User 1 can decode $\boldsymbol{X}_2$ from $\boldsymbol{X}_{1,2} \oplus (-\boldsymbol{X}_1)$; and user 2 can decode $\boldsymbol{X}_1$ from $\boldsymbol{X}_{1,2} \oplus (-\boldsymbol{X}_2)$, where $-X$ is the additive inverse of $X$.

However, when there are more than two users, this solution does not work. Consider an additional user 3 who sends linear codeword $\boldsymbol{X}_3$. The relay receives a noisy version of $\boldsymbol{X}_{1,2,3} = \boldsymbol{X}_1 + \boldsymbol{X}_2 + \boldsymbol{X}_3$. If the relay decodes and broadcasts $\boldsymbol{X}_{1,2,3}$, there is no way for any user to decode the other users' messages with $\boldsymbol{X}_{1,2,3}$ and its own message. In this case, it is not immediately obvious what the relay should do.

In this paper, we propose that the relay decodes functions of message pairs using time-division multiple-access (TDMA).

## 4.1 Coding Strategy

Let the message $W_i$ be a (binary) row vector of length $nR = k$. We define $V_{i,j} \triangleq W_i \oplus W_j$, which is also a $k$-bit row vector.

### 4.1.1 Uplink

We split the uplink transmissions into $(L-1)$ phases, each of $\frac{n}{L-1} = n'$ channel uses. In the $l$-th phase, $l \in [1, L-1]$, only users $l$ and $l+1$ transmit, i.e.,

$$X_i^{(l)} = \begin{cases} X_i(W_i), & \text{if } i = l, l+1 \\ \mathbf{0}, & \text{otherwise,} \end{cases}$$

where $X^{(l)}$ denotes $(X[(l-1)n'+1], \ldots, X[ln'])$, and $\mathbf{0}$ is the length-$n'$ all-zero row vector.

In the $l$-th transmission phase, instead of decoding messages $W_l$ and $W_{l+1}$, the relay decodes $V_{l,l+1}$ (we will explain how the relay does this using linear codes in the next section).

### 4.1.2 Downlink

Now, assuming that the relay has correctly decoded $(V_{1,2}, V_{2,3}, \ldots, V_{L-1,L})$ after $(L-1)$ transmission phases, it sends these functions back to the users.

Assuming that user $i$, $i \in [1, L]$, is able to correctly decode the functions $(V_{1,2}, V_{2,3}, \ldots, V_{L-1,L})$ sent by the relay, it performs the following (the order of decoding is important) to obtain all other users' messages:

$$W_{i+1} = V_{i,i+1} \oplus W_i \tag{7}$$
$$W_{i+2} = V_{i+1,i+2} \oplus W_{i+1} \tag{8}$$
$$\vdots$$
$$W_L = V_{L-1,L} \oplus W_{L-1} \tag{9}$$
$$W_{i-1} = V_{i-1,i} \oplus W_i \tag{10}$$
$$W_{i-2} = V_{i-2,i-1} \oplus W_{i-1} \tag{11}$$
$$\vdots$$
$$W_1 = V_{1,2} \oplus W_2. \tag{12}$$

Next, we will derive conditions on the common rate $R$ such that the relay can reliably decode $(V_{1,2}, V_{2,3}, \ldots, V_{L-1,L})$ on the uplink, and that each user can reliably decode $(V_{1,2}, V_{2,3}, \ldots, V_{L-1,L})$ on the downlink.

## 4.2 Sufficient Conditions for Reliable Uplink

For the uplink, each user $i$, $i \in [1, L]$, sends the following linear code in $n'$ channel uses:
$$X_i(W_i) = \left(W_i \odot \mathbb{G}\right) \oplus q_i, \tag{13}$$

where $\odot$ is modulo-2 vector multiplication, $\mathbb{G}$ is a fixed $k \times n'$ matrix, with each element independently and uniformly chosen over $\{0,1\}$, and $\boldsymbol{q}_i$ is a fixed row vector of length $n'$, with each element independently and uniformly chosen over $\{0,1\}$.

It can be shown that for two different messages $w_i$ and $w'_i$, their respective codewords $x_i(w_i)$ and $x_i(w'_i)$ are pair-wise independent. Using this property, Gallager showed that the codes in (13) can achieve the capacity of the binary-symmetric channel [8, page 206 (Theorem 6.2.1)].

The element-wise modulo-2 addition of the codewords of users $i$ and $j$ is given by

$$\boldsymbol{X}_i(W_i) \oplus \boldsymbol{X}_j(W_j) = \left(V_{i,j} \odot \mathbb{G}\right) \oplus \boldsymbol{q}_{i,j} \triangleq \boldsymbol{X}_{i,j}(V_{i,j}),$$

where $\boldsymbol{q}_{i,j} = \boldsymbol{q}_i \oplus \boldsymbol{q}_j$, with each element in the vector drawn according to i.i.d. uniform distribution. So, the code $\{\boldsymbol{X}_{i,j}\}$ has the same structure of that for any user $\{\boldsymbol{X}_i\}$, $\forall i \in [1, L]$.

First, consider only the $l$-th phase. We derive conditions on $R$ for *reliable* uplink. In the $l$-th phase, the uplink can be written as

$$\boldsymbol{Y}_0^{(l)} = \boldsymbol{X}_l(W_l) \oplus \boldsymbol{X}_{l+1}(W_{l+1}) \oplus \boldsymbol{E}_0^{(l)} \tag{14a}$$

$$= \boldsymbol{X}_{l,l+1}(V_{l,l+1}) \oplus \boldsymbol{E}_0^{(l)}, \tag{14b}$$

which are $n'$ independent binary-symmetric channels $X_{l,l+1} \to Y_0$, each with a cross-over probability $\rho_0$. Since the code $\{\boldsymbol{X}_{i,j}\}$ has the structure in (13), it can achieve the capacity of the binary-symmetric channel (14b), i.e., the relay can decode $V_{l,l+1}$ in the $l$-th phase with arbitrarily small error probability, if $n'$ is sufficiently large, and if

$$\frac{k}{n'} \leq C_{\text{BSC}}(\rho_0) = 1 - H(\rho_0), \tag{15}$$

where $C_{\text{BSC}}(\rho_0)$ is the capacity of the binary-symmetric channel with cross-over probability $\rho_0$ [4, page 187].

Now, we consider all $(L-1)$ phases. Since the decoding of each $V_{l,l+1}$, $l \in [1, L-1]$, only happens in one of the $(L-1)$ phases, the *effective* constraint on $R$ for reliable uplink is

$$R = \frac{k}{n} = \frac{k}{(L-1)n'} \leq \frac{C_{\text{BSC}}(\rho_0)}{L-1} = \frac{1 - H(\rho_0)}{L-1}. \tag{16}$$

### 4.3 Sufficient Conditions for Reliable Downlink

If the condition (16) is satisfied, the relay can decode $(V_{1,2}, V_{2,3}, \ldots, V_{L-1,L})$. On the downlink, the relay broadcasts $(V_{1,2}, V_{2,3}, \ldots, V_{L-1,L})$, which is a $(L-1)k$-bit concatenated message, to all users in $n$ channel uses. Since the link from the relay to any user $i$ is an independent binary-symmetric channel with cross-over probability $\rho_i$, all users can reliably decode $(V_{1,2}, V_{2,3}, \ldots, V_{L-1,L})$

if and only if

$$\frac{(L-1)k}{n} \leq C_{\text{BSC}}(\rho_i), \text{or} \tag{17}$$

$$R \leq \frac{C_{\text{BSC}}(\rho_i)}{L-1} = \frac{1-H(\rho_i)}{L-1}, \tag{18}$$

for all $i \in [1, L]$. Note that a linear code is not necessary for the downlink.

### 4.4 Achievable Rates and the Capacity

Combining (16) and (18), we have the following theorem.

**Theorem 2.** *The common-rate capacity of the binary MWRC is*

$$C = \min_{0 \leq i \leq L} \left\{ \frac{1-H(\rho_i)}{L-1} \right\} = \frac{1 - \max_{0 \leq i \leq L} H(\rho_i)}{L-1}, \tag{19}$$

*and is achievable by functional-decode-forward.*

*Proof of Theorem 2.* From Sec. 4.2, if $R \leq \frac{1-H(\rho_0)}{L-1}$, the relay is able to reliably decode $(V_{1,2}, V_{2,3}, \ldots, V_{L-1,L})$. It then broadcasts these functions to the users. From Sec. 4.3, if $R \leq \frac{1-H(\rho_i)}{L-1}$, $\forall i \in [1, L]$, all users are able to reliably decode $(V_{1,2}, V_{2,3}, \ldots, V_{L-1,L})$ from the relay. Each user can then recover $(W_1, W_2, \ldots, W_L)$. From Theorem 1, we know that this achievable common rate region (19) coincides with the upper bound to the capacity. This gives Theorem 2. □

We can also show that none of the complete-decode-forward, compress-forward, or amplify-forward coding strategies can achieve the common-rate capacity for all noise levels.

## 5 Conclusion

We have proposed a pair-wise TDMA functional-decode-forward coding strategy for the binary MWRC, and have shown that it achieves the common-rate capacity. Our proposed coding strategy can also be applied to any MWRC with an *additive* uplink, where the relay receives the summation of all user's transmit signals and noise. This includes the Gaussian MWRC, in which lattice (linear) codes can be used.